\documentclass[aps,prb,reprint,groupedaddress]{revtex4-2}

\usepackage{graphicx}

\usepackage{pifont}
\usepackage{bm}
\usepackage{xcolor}
\usepackage{color}
\usepackage[colorlinks=true,linkcolor=blue,anchorcolor=blue,citecolor=blue,urlcolor=blue]{hyperref}
\usepackage{amsmath}

\begin{document}

\title{Sequential Topological Superconductivity in a Square Lattice with Chiral Charge Density Waves}

\author{Zhong-Xian Jin$^{1,2}$}
\author{Junkang Huang$^{1,2}$}
\author{Yu-Xuan Li$^{1,2}$}
\email{yxliphy@gmail.com}
\author{Tao Zhou$^{1,2}$}
\email{tzhou@scnu.edu.cn}

\affiliation{$^1$Guangdong Basic Research Center of Excellence for Structure and Fundamental Interactions of Matter, Guangdong Provincial Key Laboratory of Quantum Engineering and Quantum Materials, School of Physics, South China Normal University, Guangzhou 510006, China\\
$^2$Guangdong-Hong Kong Joint Laboratory of Quantum Matter, Frontier Research Institute for Physics, South China Normal University, Guangzhou 510006, China
}


\begin{abstract}
The interplay between charge order and superconductivity offers a fertile ground for emergent quantum phases. Here we theoretically investigate a square-lattice superconductor coexisting with a composite charge density wave (CDW) consisting of a real bond modulation (charge bond order, CBO) and an imaginary hopping modulation (chiral flux phase, CFP) that breaks time-reversal symmetry. We uncover that, while CFP alone does not induce topology in square lattices, its coexistence with CBO drives the system into two topologically nontrivial superconducting phases with Chern numbers $C=+2$ and $C=-2$. The low-temperature thermal Hall conductivity $\kappa_{xy}$ exhibits quantized plateaus proportional to the Chern number, providing a clear experimental fingerprint. Our results establish the square lattice as a pristine platform for engineering topological superconductivity through the synergy of real and imaginary bond modulations.
\end{abstract}
\maketitle

\section{\label{Intro}Introduction}

Unconventional superconductivity rarely emerges in isolation. Instead, the microscopic coexistence and interplay of multiple order parameters are ubiquitous across various families of superconductors, including cuprates, iron-based pnictides, heavy-fermion compounds, and the recently discovered kagome superconductors and nickelate superconductors~\cite{kivelson_how_2003,Keimer2015,RevModPhys.87.457,Weng_2016,Kontani03072021,Wilson2024,10.1093/nsr/nwaf373}. This intricate competition or cooperation among distinct broken-symmetry phases profoundly dictates the physical properties of the superconducting state, making it a longstanding and central theme in modern condensed matter physics. Among these intertwined orders, charge density waves (CDWs) play a particularly prominent role.

Among the diverse spectrum of density waves, two fundamental variants are of particular theoretical and experimental significance: the chiral flux phase (CFP) and the charge bond order (CBO). On a bipartite square lattice, the CFP with a $d$-wave form factor is precisely the renowned $d$-density wave (DDW) state \cite{affleck_largen_1988,hsu_two_1991,ubbens_flux_1992,nayak_densitywave_2000,chakravarty_hidden_2001a}, which has been extensively studied in the context of cuprate pseudogap. Crucially, experimental evidence for such chiral flux phases has been reported not only in cuprates~\cite{kaminski_spontaneous_2002,mook_observation_2008,li_unusual_2008,baledent_evidence_2011,li_magnetic_2011,mangin-thro_ab_2017}, but also in iridates \cite{zhao_evidence_2016,jeong_timereversal_2017} and, more recently, in kagome superconductors $\mathrm{AV_3Sb_5}$~\cite{shumiya_intrinsic_2021,jiang_unconventional_2021,wang_electronic_2021,mielke_timereversal_2022} and FeGe~\cite{teng_discovery_2022,yin_discovery_2022,teng_magnetism_2023,han_orbital_2024}. In contrast, the CBO involves a real-part modulation of the hopping strengths. Such bond-centered charge modulations are not merely theoretical constructs but have been experimentally established across various quantum materials, including transition metal dichalcogenides (e.g., $\mathrm{TiSe_2}$~\cite{wan_nearcommensuration_2026,kim_electronically_2024} and $\mathrm{TiTe_2}$~\cite{ren_chiral_2023} and pnictides (e.g., $\mathrm{Na_2Ti_2Pn_2O}$~\cite{gillig_phononicmagnetic_2023}). Remarkably, on a square lattice, such a staggered real-part modulation naturally realizes a two-dimensional generalization of the Su-Schrieffer-Heeger (SSH) model, a paradigmatic framework for exploring topological electronic states~\cite{PhysRevA.101.063839,PhysRevA.109.022211,37gn-2g16,zbdr-v7tr,PhysRevLett.126.017601}. As highlighted above, the striking coexistence of time-reversal-breaking loop currents and structural bond modulations has recently become a focal point in the kagome family $\mathrm{AV_3Sb_5}$~\cite{ortiz_fermi_2021,luo_electronic_2022,kang_twofold_2022,fu_quantum_2021}. While these realistic materials involve intricate multi-orbital and geometric effects, exploring these distinct intertwined orders on a quintessential bipartite square lattice provides a pristine and elegant theoretical platform to unearth the universal physics driving their interplay.

While the coexistence of intertwined orders is a universal theme, its specific impact on the topological landscape of a superconductor is highly nontrivial. Recently, the interplay between time-reversal-breaking loop currents and bond modulations has been theoretically explored within the context of kagome lattices~\cite{zeng_chiralfluxphasebased_2023,zhou_topological_2023,lin_impact_2024}, revealing fascinating topological phase transitions.
In particular, Refs.~\cite{zeng_chiralfluxphasebased_2023,zhou_topological_2023} revealed that the CFP alone is sufficient to drive topological phase transitions and induce nontrivial superconducting states in kagome superconductors. 
 However, kagome systems are inherently dominated by strong geometric frustration and intricate sublattice interference, which may obscure the essential physics arising purely from the interplay of real and imaginary hopping modulations. In contrast, the bipartite square lattice offers a pristine, frustration-free platform to isolate such interplay. Importantly, previous studies on square-lattice systems have shown that the DDW order (the imaginary-part modulation) alone can already induce remarkable phenomena, including the destruction of first-order topological edge states and the emergence of higher-order topology with gapless corner states~\cite{PhysRevB.99.104517,PhysRevB.106.205128,PhysRevB.111.075169}. These findings stand in stark contrast to the kagome case, where the same imaginary modulation alone drives first-order topology, highlighting the crucial role of lattice geometry in determining the topological fate of such modulations.

Nevertheless, a crucial gap remains: all previous works on square lattices considered either the imaginary-part modulation (DDW) alone or focused on its interplay with superconductivity. The simultaneous presence of both the real-part modulation (CBO) and the imaginary-part modulation (CFP), i.e., the full CDW comprising both components, has not been systematically studied in a superconducting context. How does the coexistence of CBO and CFP reshape the topological phase diagram of a square-lattice superconductor?

In this work, we systematically investigate this generalized hybrid state combining CDW and superconductivity. We uncover a distinctive phenomenon specific to the bipartite square lattice: as the CFP strength increases in the presence of CBO, the system undergoes two successive topological phase transitions (from trivial to $C=2$, and then to $C=-2$). This sequential evolution, driven by the interplay of real and imaginary modulations, has not been previously reported in square-lattice systems. Furthermore, we establish the detailed topological phase diagram in the parameter space of chemical potential, CBO strength, and CFP strength. The chiral Majorana edge modes are directly visualized, and the thermal Hall conductivity $\kappa_{xy}$ is computed as an experimental fingerprint. Our results establish the square lattice as a pristine platform for engineering topological superconductivity through the synergy of real and imaginary bond modulations.

The remainder of this paper is organized as follows. Section II introduces the model and formalism. Section III presents the numerical results and discussion. We summarize our findings in Sec. IV.

\section{\label{sec:Model}Model and Formalism}
We consider a square lattice with coexisting superconductivity and charge density wave orders. The total Hamiltonian takes the form
\begin{eqnarray}
	H = H_{\mathrm{TB}} + H_{\mathrm{CDW}} + H_{\mathrm{SC}},
\end{eqnarray}
where $H_{\mathrm{TB}}$, $H_{\mathrm{CDW}}$, and $H_{\mathrm{SC}}$ denote the tight-binding, CDW, and superconducting pairing terms, respectively.

The tight-binding term
\begin{equation}
	H_{\mathrm{TB}} = -t \sum_{\langle i,j \rangle\sigma} c_{i\sigma}^\dagger c_{j\sigma} - \mu \sum_{i\sigma} c_{i\sigma}^\dagger c_{i\sigma}
\end{equation}
describes nearest-neighbor hopping and the chemical potential. Here $c_{i\sigma}^\dagger$ ($c_{i\sigma}$) creates (annihilates) an electron at site $i$ with spin $\sigma$, $t$ is the nearest-neighbor hopping amplitude (set to unity), and $\mu$ is the chemical potential.

\begin{figure}[tbp]
\centering
\includegraphics[width=6cm]{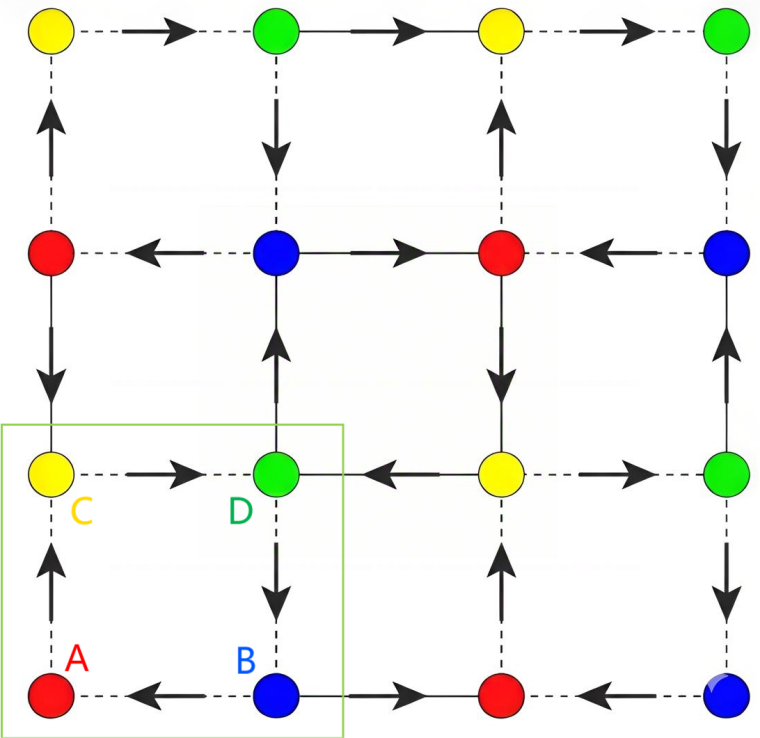}
\caption{\label{Fig1}Schematic of the real-space CDW pattern. Solid and dashed bonds represent the charge bond order (CBO), with solid (dashed) lines denoting inter-cell (intra-cell) bonds. Arrows indicate the loop currents of the chiral flux phase (CFP), with the arrow direction encoding the sign of the imaginary hopping modulation.}
\end{figure}

The charge density wave consists of two distinct orders, the CBO and the CFP, which break translational symmetry and form a $2\times2$ modulation pattern illustrated in Fig.~\ref{Fig1}. The corresponding Hamiltonian is
\begin{equation}
	H_{\mathrm{CDW}} = \lambda_{\mathrm{CBO}} \sum_{\langle i,j \rangle \sigma} \xi_{ij} c_{i\sigma}^\dagger c_{j\sigma} + i\lambda_{\mathrm{CFP}} \sum_{\langle i,j \rangle \sigma} \eta_{ij} c_{i\sigma}^\dagger c_{j\sigma} + \mathrm{H.c.},
\end{equation}
where $\lambda_{\mathrm{CBO}}$ and $\lambda_{\mathrm{CFP}}$ are the CBO and CFP strengths, respectively. The spatial modulation factors $\xi_{ij}$ and $\eta_{ij}$ are shown in Fig.~\ref{Fig1}: $\xi_{ij}=1$ ($-1$) on inter-cell (intra-cell) bonds, denoted by solid (dashed) lines; and $\eta_{ij}=1$ ($-1$) when the bond is aligned (opposite) to the arrow direction.

The superconducting pairing is assumed to be an isotropic $s$-wave form. 
In momentum space, the pairing term can be written as
\begin{equation}
	H_{\mathrm{SC}} =
	\sum_{\mathbf{k}, \alpha}
	\left(
	\Delta\, c^\dagger_{\mathbf{k}\alpha\uparrow}
	c^\dagger_{-\mathbf{k}\alpha\downarrow}
	+ \text{H.c.}
	\right),
\end{equation}
where $\alpha$ labels the sublattice degree of freedom, and $\Delta$ denotes a uniform $s$-wave pairing amplitude.
 
 Fourier transforming to momentum space gives
 \begin{eqnarray}
 	H &=& -t \sum_{\mathbf{k}\langle\alpha\alpha'\rangle\sigma} \left[ c_{\mathbf{k}\alpha\sigma}^\dagger c_{\mathbf{k}\alpha'\sigma} e^{-i\mathbf{k} \cdot (\mathbf{r}_\alpha - \mathbf{r}_{\alpha'})} + \text{H.c.}\right] \nonumber \\
 	&&+ \lambda_{\mathrm{CBO}} \sum_{\mathbf{k}\langle\alpha\alpha'\rangle\sigma} \left[ \xi_{\alpha\alpha'} c_{\mathbf{k}\alpha\sigma}^\dagger c_{\mathbf{k}\alpha'\sigma} e^{-i\mathbf{k} \cdot (\mathbf{r}_\alpha - \mathbf{r}_{\alpha'})} + \text{H.c.} \right] \nonumber \\
 	&&+ i\lambda_{\mathrm{CFP}} \sum_{\mathbf{k}\langle\alpha\alpha'\rangle\sigma} \left[ \eta_{\alpha\alpha'} c_{\mathbf{k}\alpha\sigma}^\dagger c_{\mathbf{k}\alpha'\sigma} e^{-i\mathbf{k} \cdot (\mathbf{r}_\alpha - \mathbf{r}_{\alpha'})} + \text{H.c.} \right] \nonumber\\
 	&&+ \sum_{\mathbf{k}\alpha} (\Delta c_{\mathbf{k}\alpha\uparrow}^\dagger c_{-\mathbf{k}\alpha\downarrow}^\dagger + \text{H.c.}) - \mu \sum_{\mathbf{k}\alpha\sigma} c_{\mathbf{k}\alpha\sigma}^\dagger c_{\mathbf{k}\alpha\sigma},
 \end{eqnarray}
 where $\alpha,\alpha'$ label the sublattice degrees of freedom within the enlarged unit cell, and $\mathbf{r}_\alpha$ are their positions.
 
 This Hamiltonian belongs to class D of the Altland-Zirnbauer classification \cite{schnyder_classification_2008}, where time-reversal symmetry is broken while particle-hole symmetry is preserved. Consequently, the topological properties of the system are characterized by the Chern number. For two-dimensional class-D systems without TRS, the Chern number can take non-zero values, which directly gives rise to chiral Majorana edge modes.

To study the edge states, we impose periodic boundary conditions along the $x$ direction and open boundary conditions along the $y$ direction. This strip geometry is adopted to visualize chiral edge modes localized at the upper and lower boundaries along the $y$ direction. The Hamiltonian then takes the form $H = \sum_{y,k_x} \psi^{\dagger}(y,k_x) M(k_x) \psi(y,k_x)$, where $M(k_x)$ is an $8N_y \times 8N_y$ matrix and the basis is $\psi(y,k_x)^{\dagger} = (c_{y,k_xA}^{\dagger}, c_{y,k_xB}^{\dagger}, c_{y,k_xC}^{\dagger}, c_{y,k_xD}^{\dagger}, c_{y,k_xA}, c_{y,k_xB}, c_{y,k_xC}, c_{y,k_xD})$ for $y = 1, 2, \dots, N_y$, with $N_y$ the number of unit cells along $y$. Diagonalizing $M(k_x)$ yields the energy spectrum as a function of $k_x$; gapless states inside the bulk gap signal topological edge modes.

To make contact with possible experiments, we evaluate the thermal Hall conductivity $\kappa_{xy}$. As Majorana quasiparticles are neutral, they produce no charge Hall response and can only be probed via thermal Hall transport. This quantity can be computed from the linear-response Kubo formula \cite{mo_coexistence_2025,qin_energy_2011}:
\begin{equation}
	\kappa_{xy} =
	-\frac{k_B^2 T}{2\hbar}
	\int_{-\infty}^{+\infty} dE \,
	\frac{E^2}{(k_B T)^2}
	\sigma(E)\, f'(E),
\end{equation}
where $f(E) = [\exp(E/k_B T)+1]^{-1}$ is the Fermi-Dirac distribution, and $\sigma(E)$ denotes the integrated Berry curvature below energy $E$. 

In the zero-temperature limit, $-f'(E)$ approaches a delta function centered at the Fermi level, and the above expression reduces to
\begin{equation}
	\kappa_{xy} = C \kappa_0,
\end{equation}
where $C$ is the Chern number and $\kappa_0 = (\pi k_B^2 / 12\hbar)\, T$ is the thermal conductance quantum carried by a single chiral Majorana edge mode \cite{yoshida_observation_2025a}.

To map out the locations of gapless states in the Brillouin zone and thereby identify the Dirac cones and band-touching points associated with the topological phase transitions, we compute the spectral function $A(\mathbf{k},\omega)$. Diagonalizing the BdG Hamiltonian yields eigenvalues $E_n(\mathbf{k})$ and eigenstates with sublattice-resolved components $u_{jn}(\mathbf{k})$ and $v_{jn}(\mathbf{k})$ in the particle-hole space, where $j$ labels the four sublattice sites within the enlarged unit cell. The spectral function is given by

\begin{equation}
	\begin{split}
		A(\mathbf{k}, \omega) = &-\frac{1}{\pi} \mathrm{Im} \sum_{j=1}^{4} \sum_{n}
		\left[
		\frac{|u_{jn}(\mathbf{k})|^2}{\omega - E_n(\mathbf{k}) + i\Gamma} \right. \\
		&\left. + \frac{|v_{jn}(\mathbf{k})|^2}{\omega + E_n(\mathbf{k}) + i\Gamma}
		\right],
	\end{split}
\end{equation}
where $\Gamma = 0.01$ is a small broadening factor introduced to avoid singularities. The zero-energy spectral function $A(\mathbf{k},0)$ directly reveals the momentum-space locations of gapless excitations, which will be used below to identify the Dirac cones and band-touching points characteristic of the topological phase transitions.

In our calculations, we set the $s$-wave pairing amplitude to $\Delta = 0.2$, and treat $\mu$, $\lambda_{\mathrm{CBO}}$, and $\lambda_{\mathrm{CFP}}$ as tunable parameters to map out the topological phase diagram. The CDW pattern illustrated in Fig.~\ref{Fig1} has a period of $2a \times 2a$, where $a$ is the nearest-neighbor distance; we take $2a$ as the length unit.

\section{Numerical results and discussion}

\begin{figure}[tbp]
	\centering
	\includegraphics[width=1.0\linewidth]{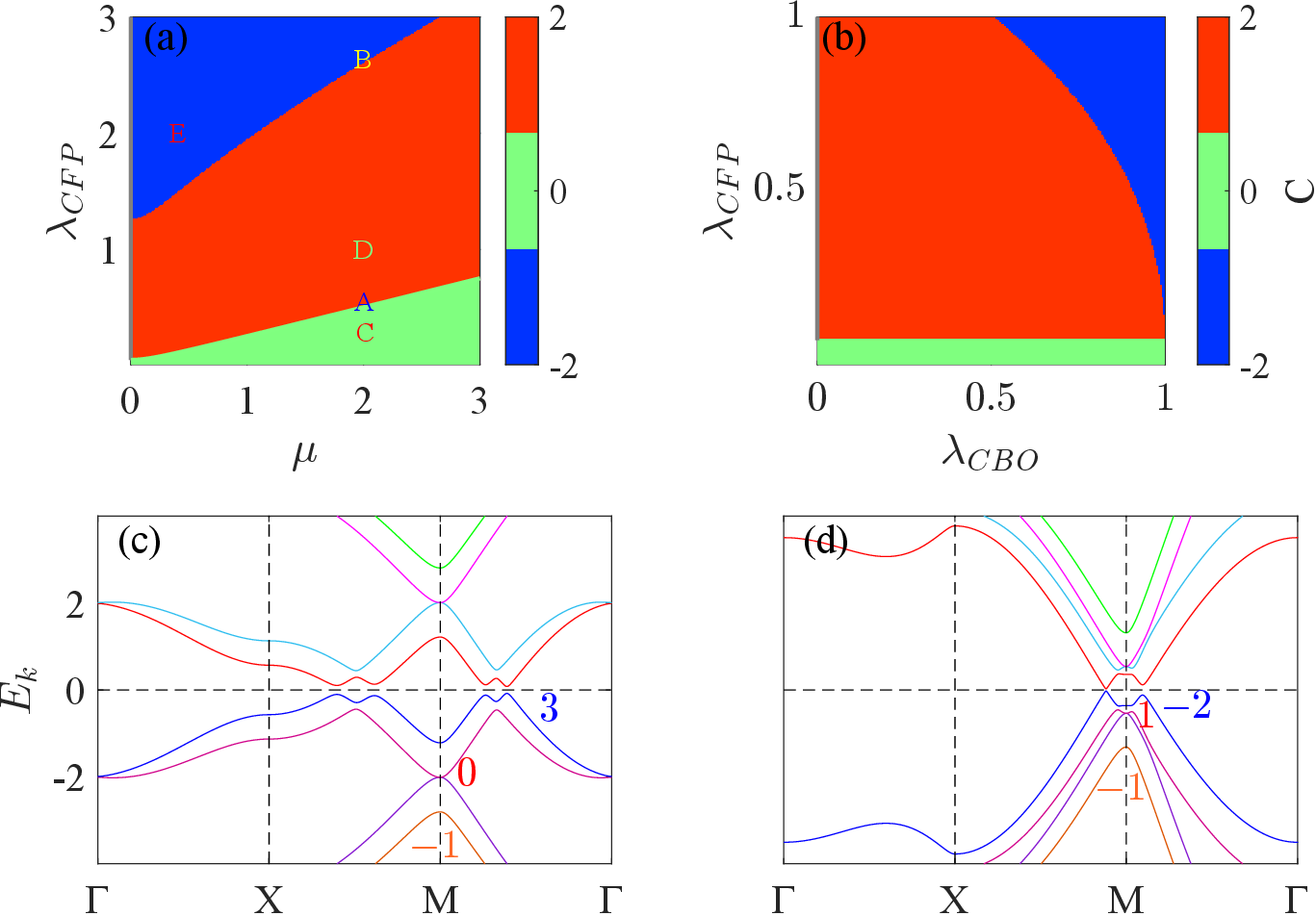}
	\caption{\label{Fig2}Topological phase diagrams and band-resolved Chern numbers. (a) Phase diagram in the $\mu$-$\lambda_{\mathrm{CFP}}$ plane with $\lambda_{\mathrm{CBO}}=0.2$. (b) Phase diagram in the $\lambda_{\mathrm{CBO}}$-$\lambda_{\mathrm{CFP}}$ plane with $\mu=0.2$. (c),(d) Band structures at points D and E in (a), corresponding to the $C=2$ and $C=-2$ phases with $(\mu,\lambda_{\mathrm{CFP}})=(2,1)$ and $(0.5,2)$, respectively. The Chern number of each occupied band is labeled in each panel.}
\end{figure}

We first construct the topological phase diagrams in the $\mu$-$\lambda_{\mathrm{CFP}}$ and $\lambda_{\mathrm{CBO}}$-$\lambda_{\mathrm{CFP}}$ planes, shown in Figs.~\ref{Fig2}(a) and (b), respectively. Three phases with Chern numbers $C=-2$, $0$, and $2$ are identified, separated by two phase boundaries: one between $C=0$ and $C=2$, and another between $C=2$ and $C=-2$. The phase diagram exhibits the following symmetry properties: reversing the sign of $\mu$ flips the Chern number, while reversing $\lambda_{\mathrm{CBO}}$ leaves it unchanged; reversing $\lambda_{\mathrm{CFP}}$ also flips the Chern number. These symmetry properties reflect the underlying particle-hole symmetry of the BdG Hamiltonian: $\mu \to -\mu$ corresponds to the particle-hole conjugation of the BdG spectrum, and the sign change of $\lambda_{\mathrm{CFP}}$ reverses the chirality of the loop currents. Consequently, it suffices to present the phase diagrams for nonnegative values of $\mu$, $\lambda_{\mathrm{CBO}}$, and $\lambda_{\mathrm{CFP}}$.

In Fig.~\ref{Fig2}(a), where $\lambda_{\mathrm{CBO}}$ is fixed, the critical $\lambda_{\mathrm{CFP}}$ for both phase transitions increases monotonically with $\mu$. This indicates that a larger chemical potential requires a stronger CFP to close and reopen the bulk gap.

In Fig.~\ref{Fig2}(b), where $\mu$ is fixed, the behavior of the two transitions differs qualitatively. The first transition, from $C=0$ to $C=2$, occurs at a critical $\lambda_{\mathrm{CFP}}$ that is independent of $\lambda_{\mathrm{CBO}}$, as reflected by the horizontal phase boundary. However, this transition requires a finite $\lambda_{\mathrm{CBO}}$: at $\lambda_{\mathrm{CBO}}=0$, the gap closes at the would-be transition point but does not reopen beyond it, so the system remains gapless and no topological transition occurs. The second transition, from $C=2$ to $C=-2$, exhibits a different trend: its critical $\lambda_{\mathrm{CFP}}$ decreases monotonically as $\lambda_{\mathrm{CBO}}$ increases.

To further elucidate the topological nature of the two nontrivial phases, we compute the Chern numbers of individual occupied bands at representative points in the phase diagram, as shown in Figs.~\ref{Fig2}(c) and (d) for the $C=2$ and $C=-2$ phases, respectively. For the $C=2$ phase [Fig.~\ref{Fig2}(c)], the composite band formed by the second and third bands carries $C=0$; the total Chern number therefore comes entirely from the first ($C=-1$) and fourth ($C=3$) bands. For the $C=-2$ phase [Fig.~\ref{Fig2}(d)], the composite band acquires $C=1$, and the total Chern number $C=-2$ is contributed by the first ($C=-1$), fourth ($C=-2$), and composite ($C=1$) bands.

\begin{figure}[tbp]
	\centering
	\includegraphics[width=\linewidth]{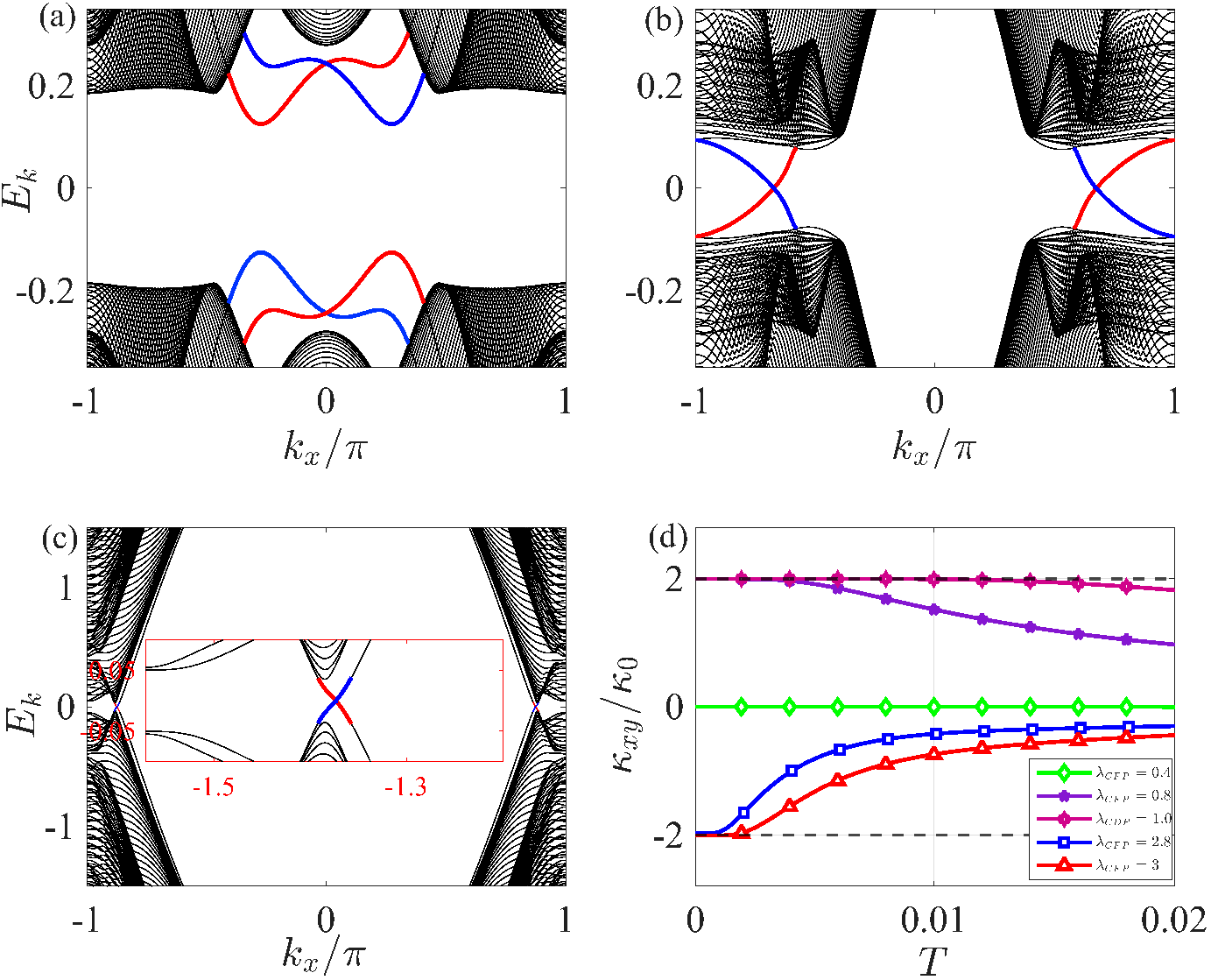}
	\caption{\label{Fig3} Energy spectra under partial open boundary conditions and thermal Hall conductivity. (a)--(c) Energy spectra with $\lambda_{\mathrm{CBO}}=0.2$ at points C, D, and E in Fig.~\ref{Fig2}(a), with $(\mu,\lambda_{\mathrm{CFP}})=(2,0.4)$, $(2,1)$, and $(0.5,2)$, respectively. (d) Thermal Hall conductivity $\kappa_{xy}$ versus $T$ for $(\mu,\lambda_{\mathrm{CBO}})=(2,0.2)$ at several $\lambda_{\mathrm{CFP}}$ values.}
\end{figure}

According to the bulk-edge correspondence, a topological phase with Chern number $C$ supports $|C|$ chiral edge modes in a cylinder geometry. We therefore examine the edge-state spectra under partial open boundary conditions for the three phases identified in Fig.~\ref{Fig2}. Figures~\ref{Fig3}(a)-\ref{Fig3}(c) show the energy spectra for the $C=0$, $2$, and $-2$ phases, respectively. In the trivial phase [Fig.~\ref{Fig3}(a)], the edge bands are fully gapped with no in-gap states. For $C=2$ [Fig.~\ref{Fig3}(b)], edge bands localized at $y=1$ (red) and $y=N_y$ (blue) are visible, with two pairs of chiral edge modes crossing at $k_x = \pm 0.67\pi$, consistent with the Chern number. For $C=-2$ [Fig.~\ref{Fig3}(c)], the gap is extremely narrow despite $\Delta=0.2$, owing to CFP-induced suppression of superconductivity. The inset resolves two pairs of chiral edge modes at $k_x = \pm 0.87\pi$, confirming the $C=-2$ topology.

For a class-D topological superconductor, the zero-temperature thermal Hall conductivity is quantized as given in Eq.~(7). This provides a direct experimental probe of the topological invariant. Figure~\ref{Fig3}(d) shows $\kappa_{xy}/\kappa_0$ as a function of $T$ for several $\lambda_{\mathrm{CFP}}$ values. At low $T$, all curves approach the quantized values $C\kappa_0$ corresponding to their respective Chern numbers. As $T$ increases, $\kappa_{xy}/\kappa_0$ decays toward zero, with the onset temperature determined by the bulk energy gap $\Delta E$: smaller gaps lead to earlier deviations from quantization. For $\lambda_{\mathrm{CFP}}=0.8, 1.0, 2.8, 3.0$, we find $\Delta E = 0.023, 0.080, 0.005, 0.009$, respectively. The extremely small gaps for $\lambda_{\mathrm{CFP}}=2.8$ and $3.0$ suppress the quantized plateau even at very low temperatures, while for $\lambda_{\mathrm{CFP}}=0.4$, the large gap maintains a flat plateau up to $T\approx 0.02$. These results demonstrate that thermal Hall conductivity measurements can serve as a viable tool for detecting the topological invariant and mapping out the phase diagram in future experiments.

We now turn to the microscopic origin of the sequential topological transitions observed in the phase diagram. Since the system has uniform $s$-wave pairing, the closing and reopening of the bulk BdG gap are directly tied to the normal-state electronic structure near the Fermi level. Thus, examining the normal-state band reconstruction induced by CBO and CFP offers a natural route to understanding the underlying mechanism.

\begin{figure}[tbp]
	\centering
	\includegraphics[width=1\linewidth]{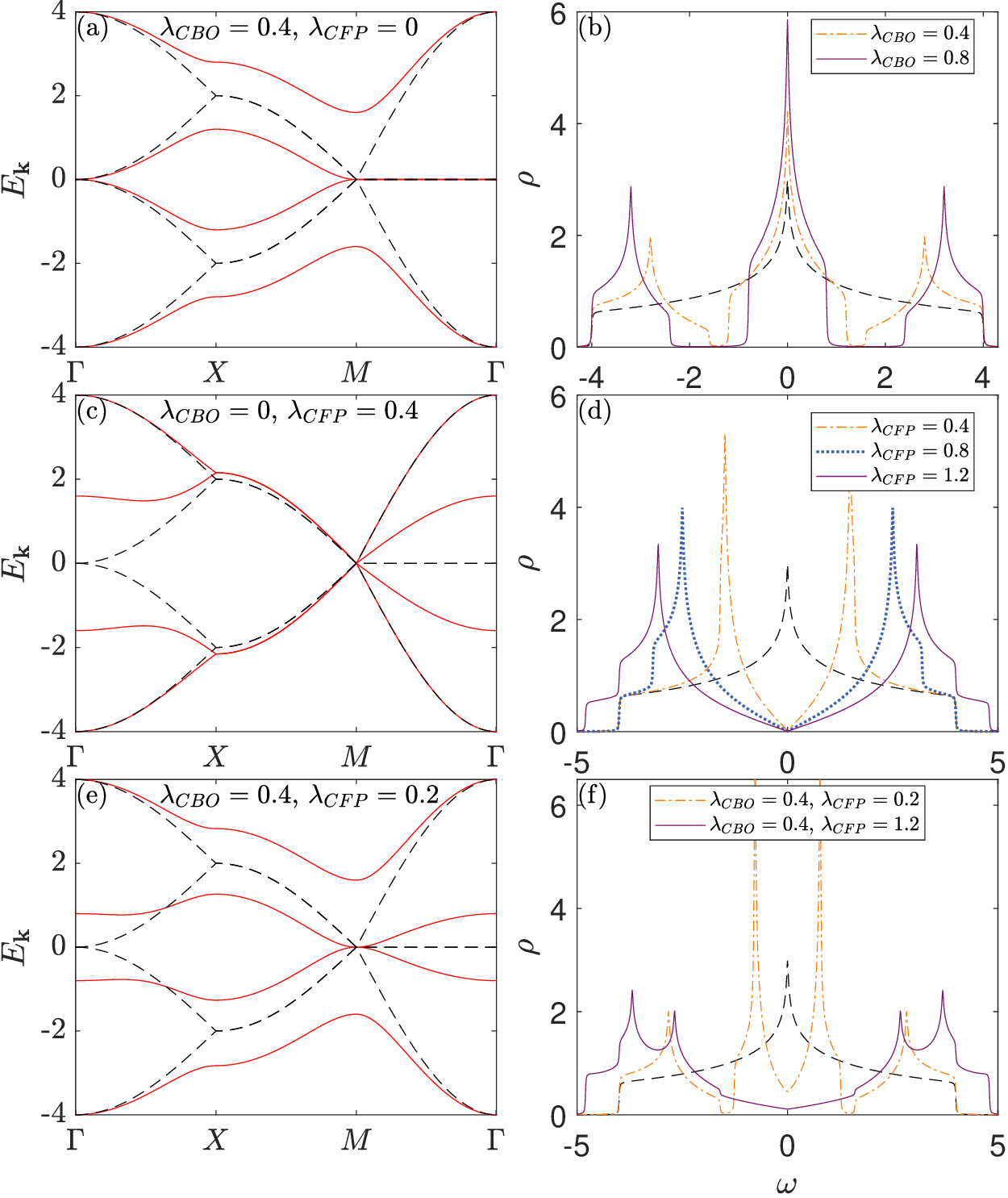}
	\caption{\label{Fig4}Energy bands (left panels) and DOS (right panels) along the high-symmetry path $\Gamma$-X-M-$\Gamma$. (a),(b) CBO only ($\lambda_{\mathrm{CFP}}=0$); (c),(d) CFP only ($\lambda_{\mathrm{CBO}}=0$); (e),(f) both CBO and CFP. Dashed curves indicate the corresponding quantities without CDW.}
\end{figure}

We first consider the case with CBO only, shown in Fig.~\ref{Fig4}(a) for $(\lambda_{\mathrm{CBO}},\lambda_{\mathrm{CFP}})=(0.4,0)$. The real-space modulation of hopping amplitudes induces a pronounced anisotropic band reconstruction. In particular, the low-energy bands remain gapless along the M--$\Gamma$ direction, while the higher-energy bands are fully split into separated branches. This reconstruction produces a sharp zero-energy peak in the density of states (DOS), accompanied by two distinct gapped features at higher energies [Fig.~\ref{Fig4}(b)].

We next consider the case with only CFP. The imaginary hopping modulation predominantly affects low-energy states near the Fermi level. As shown in the Appendix, the CFP contribution vanishes identically at the M point. Consequently, the CFP term alone cannot lift the symmetry-protected degeneracy at M, leaving the second and third bands degenerate. This persistent degeneracy prevents assigning well-defined Chern numbers to these two bands individually; they must be treated as a single composite band. The preserved band crossing at M [Fig.~\ref{Fig4}(c)] yields a characteristic V-shaped DOS near zero energy [Fig.~\ref{Fig4}(d)]. This symmetry constraint plays a key role not only in the topological classification of the reconstructed bands but also in the gap-closing processes associated with the phase transitions.

We now turn to the coexistence of CBO and CFP. For small $\lambda_{\mathrm{CFP}}$, the CFP acts as a weak perturbation that primarily reshapes the low-energy dispersion. The degeneracy at M remains intact because the CFP contribution vanishes at this point, while the CFP opens gaps in the low-energy bands away from M. In this regime, the two orders dominate distinct energy regimes: CBO splits the high-energy bands, while CFP modifies the low-energy states near the Fermi level. As a result, the CFP correction to the already separated high-energy bands is negligible [Fig.~\ref{Fig4}(e)], and the total DOS can be approximated as a superposition of the individual contributions [Fig.~\ref{Fig4}(f)].

For sufficiently large $\lambda_{\mathrm{CFP}}$, the reconstruction of low-energy bands induced by CFP becomes substantial and can no longer be treated as a weak perturbation. In this strong-CFP regime, the real bond modulation from CBO and the imaginary hopping modulation from CFP act cooperatively to reshape the low-energy effective Hamiltonian: they simultaneously modify the momentum-dependent dispersion from CBO and the time-reversal-breaking contribution from CFP. As a result, the normal-state electronic spectrum undergoes significant reconstruction, manifested as the splitting of CFP-related DOS peaks in the coexistence regime [Fig.~\ref{Fig4}(f)].
We note that the strong renormalization of the low-energy bands by CFP also suppresses the superconducting pairing: it modifies the DOS at the Fermi level and introduces additional quasiparticle scattering that reduces the effective pairing strength, leading to a remarkably narrow bulk gap even for a fixed bare pairing amplitude $\Delta$. This accounts for the extremely small gap observed in the $C=-2$ phase in Fig.~\ref{Fig3}(c).

The evolution of the low-energy electronic structure provides the microscopic mechanism by which CFP drives the system through successive symmetry-constrained gap-closing points. Specifically, the CFP-induced renormalization of the low-energy effective BdG parameters controls the closing and reopening of the bulk gap. The two sequential topological transitions are associated with distinct gap-closing scenarios, rooted in different symmetry constraints at different momenta: the first transition proceeds via quadratic band touching at the $\Gamma$ point protected by the $P$ symmetry, while the second transition arises from linear Dirac crossings along the $\mathrm{X}$--$\mathrm{M}$ lines where the symmetry constraint is lifted. This symmetry-driven distinction will be further elaborated via $k\cdot p$ perturbation theory in the following section.

The symmetry-enforced band degeneracy at the M point also imposes a fundamental constraint on the topological characterization of the system. Since the two degenerate bands remain glued together at this high-symmetry point, their topological contribution cannot be assigned to each band individually, and must instead be evaluated by treating the composite band as a single connected manifold.

An analogous symmetry constraint governs the gap-closing behavior at the $\Gamma$ point, which accounts for the distinct feature of the first phase boundary. Within the $k\cdot p$ low-energy expansion, the CBO term makes no contribution to the zeroth-order Hamiltonian at the $\Gamma$ point. Furthermore, although all linear-in-$k$ terms originate entirely from the CBO, they vanish identically when projected onto the zero-energy subspace. Accordingly, the critical condition for gap closing at this high-symmetry point is independent of $\lambda_{\mathrm{CBO}}$, which naturally explains the horizontal phase boundary of the $C=0\to C=2$ transition in Fig.~\ref{Fig2}(b).

Turning to the chemical potential dependence of the phase boundaries, Fig.~\ref{Fig2}(a) shows that the critical $\lambda_{\mathrm{CFP}}$ for both topological transitions increases monotonically with $\mu$. This behavior follows directly from the energy-resolved effect of the CFP order: CFP predominantly reconstructs low-energy states near the Fermi level. When $\mu$ is shifted away from this CFP-sensitive low-energy regime, stronger CFP-induced renormalization is required to bring the reconstructed bands into the gap-closing regime of the BdG spectrum. The critical CFP strength therefore increases systematically with increasing $\mu$.

Finally, the spectral features identified above provide experimentally accessible signatures for the coexistence of CBO and CFP orders. Specifically, the zero-energy DOS peak is a fingerprint of CBO-dominated band reconstruction, the V-shaped DOS structure near zero energy is characteristic of CFP order, and the peak splitting in the DOS signals the strong cooperative effect of the two orders. These spectral features can in principle be detected by angle-resolved photoemission spectroscopy and low-temperature scanning tunneling spectroscopy, offering measurable probes for identifying the interplay between different charge density wave components.

\begin{figure}[tbp]
	\centering
	\includegraphics[width=1\linewidth]{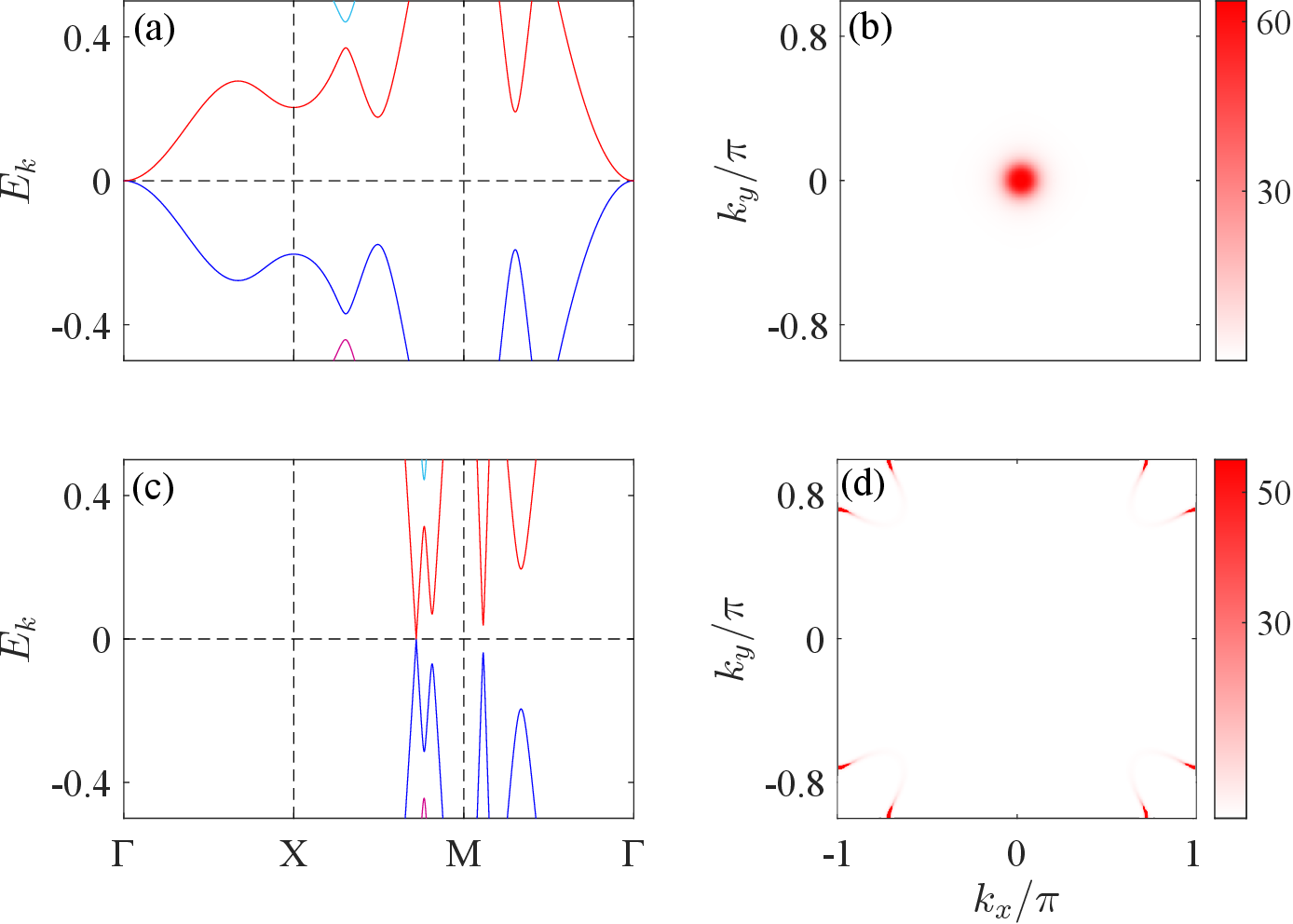}	
	\caption{(a),(b) Band structure and zero-energy spectral function at transition point A in Fig.~\ref{Fig2}(a), with $(\mu,\lambda_{\mathrm{CFP}})=(2,0.5025)$. (c),(d) Band structure and zero-energy spectral function at transition point B in Fig.~\ref{Fig2}(a), with $(\mu,\lambda_{\mathrm{CFP}})=(2,2.56)$.}
	\label{Fig5}
\end{figure}

The sequential transitions between distinct Chern phases observed in the phase diagram originate from distinct bulk gap-closing processes. Since the Chern number is a bulk topological invariant, it can change only when the quasiparticle gap closes and reopens \cite{PhysRevLett.49.405,RevModPhys.82.3045}. The change in the Chern number across a transition is determined by the number and low-energy dispersion of the gap-closing points \cite{PhysRevLett.61.2015,RevModPhys.83.1057,10.1093/acprof:oso/9780199564842.001.0001}: a quadratic band-touching point contributes $\Delta C = \pm 2$, while each linear Dirac cone contributes $\Delta C = \pm 1$. To reveal which mechanism governs each transition, we examine in Fig.~\ref{Fig5} the band structures at the two phase boundaries identified in Fig.~\ref{Fig2}(a).

In Fig.~\ref{Fig5}(a), the bands close at $\Gamma$ on the phase boundary between $C=0$ and $C=2$, where a pronounced bright feature appears in the zero-energy spectral function at the $\Gamma$ point [Fig.~\ref{Fig5}(b)]. To characterize this gap-closing behavior, we perform a $k \cdot p$ expansion of the BdG Hamiltonian \cite{Luttinger_Motion_1955,Winkler_Spin–Orbit_2003,Barry_Beyond_2016}.
\begin{equation}
	H(\mathbf{k}) = H_{\Gamma}(0) + k_x V_x + k_y V_y + \mathcal{O}(k^2),
\end{equation}
where $V_i = \partial_{k_i} H(\mathbf{k})|_{\mathbf{k}=0}$ are velocity operators encoding the linear momentum coupling, and $\mathcal{O}(k^2)$ contains higher-order terms.

Introducing the symmetry operator $P = \sigma_x s_x \tau_0$, one finds
\begin{equation}
	[H_{\Gamma}(0),P] = 0, \qquad \{P,V_i\} = 0.
\end{equation}
As shown in Appendix A, all gap-closing states at $\Gamma$ share the same $P$ parity. We therefore define the projection onto the zero-energy subspace as $\Pi = \sum_{\alpha} |u_\alpha\rangle \langle u_\alpha|$, with $H_{\Gamma}(0)|u_\alpha\rangle = E_0 |u_\alpha\rangle$.

Within the zero-energy subspace, degenerate $k\cdot p$ perturbation theory gives the first-order effective Hamiltonian
\begin{equation}
	H_{\mathrm{eff}}^{(1)} = \Pi (k_x V_x + k_y V_y)\Pi.
\end{equation}
Using $\{P,V_i\}=0$ and the fact that all states in this subspace share the same $P$ parity, we obtain
\begin{equation}
	\Pi V_x \Pi = \Pi V_y \Pi = 0,
\end{equation}
and hence $H_{\mathrm{eff}}^{(1)}=0$. The leading term therefore appears at second order (a nonzero second-order contribution is shown in the Appendix). Consequently, the gap-closing at $\Gamma$ exhibits quadratic band touching in all momentum directions, yielding a Chern number change $\Delta C = \pm 2$. An analytical solution at $\Gamma$ gives the phase boundary $\lambda_{\mathrm{CFP}} = \sqrt{\Delta^2 + \mu^2}/4$, independent of $\lambda_{\mathrm{CBO}}$ and consistent with the horizontal boundary in Fig.~\ref{Fig2}(b).

Along the phase boundary between $C=2$ and $C=-2$ [Fig.~\ref{Fig5}(c)], the bulk gap closes along the Brillouin-zone boundary X--M. Here the zeroth-order Hamiltonian no longer commutes with $P=\sigma_x s_x \tau_0$, so the symmetry constraint on the first-order terms is lifted. As a result, linear momentum-dependent terms become allowed in the effective Hamiltonian, giving rise to linearly dispersing Dirac cones. Eight such cones appear along the Brillouin-zone boundary [Fig.~\ref{Fig5}(d)], among which only four are inequivalent. This yields a Chern number change $\Delta C = \pm 4$.

\section{summary}
In summary, we have presented a comprehensive study of intertwined topological phases in a square-lattice superconductor hosting coexisting charge bond order (CBO) and chiral flux phase (CFP). By utilizing a clean bipartite square lattice, we isolate the fundamental physics from the intricate geometric complications inherent to multi-orbital or Kagome systems. Our key finding is the discovery of a sequential topological phase transition driven by tuning the CFP strength. We map out the detailed topological phase diagram and demonstrate the bulk-boundary correspondence through the direct visualization of robust chiral edge modes. Furthermore, we propose that the thermal Hall conductivity can serve as an experimental signature to identify these exotic phases. This work not only advances our understanding of topological superconductivity in the presence of competing and intertwined orders but also provides a concrete guide for designing topological quantum states in square-lattice materials and heterostructures.

\appendix
\section{k$\cdot$p Expansion and Quadratic Band Touching at $\Gamma$}

In the Nambu basis, the BdG Hamiltonian is
\begin{equation}
	\begin{aligned}
	H(\mathbf{k})&=
	-2t\left\{ \cos\left(\frac{k_x}{2}\right)\sigma_0 s_x - \cos\left(\frac{k_y}{2}\right)\sigma_x s_0 \right\}\tau_z \\
	&-2\lambda_{\mathrm{CBO}}\left\{\sin\left(\frac{k_x}{2}\right)\sigma_0 s_y + \sin\left(\frac{k_y}{2}\right)\sigma_y s_0\right\}\tau_z \\
	&-2\lambda_{\mathrm{CFP}}\left\{\cos\left(\frac{k_x}{2}\right)\sigma_z s_y - \cos\left(\frac{k_y}{2}\right)\sigma_y s_z\right\}\tau_0 \\
	&+ \Delta_0 \sigma_0 s_0 \tau_x-\mu\sigma_0 s_0 \tau_z,
	\end{aligned}
\end{equation}
where $\sigma_i$, $s_i$, and $\tau_i$ are Pauli matrices acting on the orbital, spin, and particle-hole degrees of freedom, respectively.It follows that the Hamiltonian term of the CFP vanishes identically at the \(\mathrm{M}\) point.

Near the $\Gamma$ point, we expand the lattice Hamiltonian using $\sin k_i \approx k_i$ and $\cos k_i \approx 1 - k_i^2/2$. The $k\cdot p$ expansion gives
\begin{equation}
	H(\mathbf{k}) = H_0 + \frac{\hbar}{m}\mathbf{k}\cdot\mathbf{p} + \frac{\hbar^2 k^2}{2m},
\end{equation}
where $H_0$ is the Hamiltonian at the expansion point, the second term is the leading linear correction, and the last term is the free-electron quadratic contribution.

Equivalently, a Taylor expansion around $\Gamma$ gives
\begin{equation}
	\begin{aligned}
		H(\mathbf{k}) &= H_{\Gamma}(0) + \sum_i k_i\left.\frac{\partial H}{\partial k_i}\right|_{\Gamma} \\
		&\quad + \frac{1}{2}\sum_{ij} k_i k_j\left.\frac{\partial^2 H}{\partial k_i \partial k_j}\right|_{\Gamma}.
	\end{aligned}
\end{equation}

The unperturbed Hamiltonian at $\Gamma$ is
\begin{equation}
	\begin{aligned}
 	 H_{\Gamma}(0) &=
	  -2t(\sigma_0 s_x + \sigma_x s_0)\tau_z \\
	 &\quad-2\lambda_{\mathrm{CFP}}(\sigma_z s_y - \sigma_y s_z)\tau_0\\
	 &\quad+\Delta_0 \sigma_0 s_0 \tau_x- \mu\sigma_0 s_0 \tau_z .
	\end{aligned}
\end{equation}
The linear momentum dependence is $H^{(1)}(\mathbf{k}) = k_x V_x + k_y V_y$, with
\begin{equation}
	\begin{aligned}
	V_x &= \left.\frac{\partial H}{\partial k_x}\right|_{\Gamma}
	= -\lambda_{\mathrm{CBO}}\,\sigma_0 s_y \tau_z,\\
	V_y &= \left.\frac{\partial H}{\partial k_y}\right|_{\Gamma}
	= -\lambda_{\mathrm{CBO}}\,\sigma_y s_0 \tau_z.
	\end{aligned}
\end{equation}
The second-order contributions are
\begin{equation}
	H^{(2)}(\mathbf{k}) =
	\frac{1}{2}k_x^2 W_{xx}
	+
	\frac{1}{2}k_y^2 W_{yy}
	+
	k_x k_y W_{xy},
\end{equation}
with
\begin{equation}
	\begin{aligned}
		W_{xy}&=0,\\
		W_{xx}=\lambda_{\mathrm{CFP}}\sigma_z s_y \tau_0,& \quad
		W_{yy}=\lambda_{\mathrm{CFP}}\sigma_y s_z \tau_0.
	\end{aligned}
\end{equation}
\vspace{6pt}

Consider the unitary Hermitian operator
\begin{equation}
	P = \sigma_x s_x \tau_0, \qquad P^2=1.
\end{equation}
It satisfies
\begin{equation}
	\begin{aligned}
	[H_{\Gamma}(0),P]&=0,\\  \{P,V_i\}=0, \qquad [&P,W_{ij}]=0.
	\end{aligned}
\end{equation}
We project onto the zero-energy subspace via
\begin{equation}
	\Pi = \sum_{\alpha} |u_{\alpha}\rangle \langle u_{\alpha}|,
	\qquad H(0)|u_{\alpha}\rangle = E_0 |u_{\alpha}\rangle.
\end{equation}
The effective Hamiltonians are
\begin{equation}
	H_{\mathrm{eff}}^{(1)} = \Pi (k_x V_x + k_y V_y)\Pi
\end{equation}
\begin{equation}
	H_{\mathrm{eff}}^{(2)} =
	\Pi \left(
	\frac{1}{2}k_x^2 W_{xx}
	+
	\frac{1}{2}k_y^2 W_{yy}
	+
	k_x k_y W_{xy}
	\right)\Pi
\end{equation}

Since $[H_{\Gamma}(0),P]=0$, we may choose the zero-energy eigenstates as eigenstates of $P$,
\begin{equation}
	P|u_{\alpha}\rangle = p_{\alpha}|u_{\alpha}\rangle, \qquad p_{\alpha}=\pm1.
\end{equation}
Within this subspace, matrix elements satisfy
\begin{equation}
	\begin{aligned}
	\langle u_{\alpha}|V_i|u_{\beta}\rangle
	&= -p_{\alpha}p_{\beta}
	\langle u_{\alpha}|V_i|u_{\beta}\rangle,
    \\
	\langle u_{\alpha}|W_{ij}|u_{\beta}\rangle
	&= p_{\alpha}p_{\beta}
	\langle u_{\alpha}|W_{ij}|u_{\beta}\rangle.
	\end{aligned}
\end{equation}

All gap-closing states belong to the same symmetry sector (numerically verified), so $p_{\alpha}p_{\beta}=1$. Hence
\begin{equation}
	\langle u_{\alpha}|V_i|u_{\beta}\rangle = 0,
	\qquad
	\langle u_{\alpha}|W_{ij}|u_{\beta}\rangle \neq 0,
\end{equation}
and therefore
\begin{equation}
	\Pi V_i \Pi = 0,
	\qquad
	\Pi W_{ij} \Pi \neq 0.
\end{equation}
Thus $H_{\mathrm{eff}}^{(1)}=0$ while $H_{\mathrm{eff}}^{(2)}$ is finite. The leading momentum dependence is therefore quadratic, implying quadratic band touching at $\Gamma$ in all momentum directions.

\bibliography{DDW.bib}

\end{document}